\documentclass[pdflatex,sn-mathphys-num]{sn-jnl}

\usepackage[mathlines]{lineno} 
\usepackage{graphicx}%
\usepackage{multirow}%
\usepackage{mathtools}
\usepackage{amsmath,amssymb,amsfonts}%
\usepackage{amsthm}%
\usepackage{mathrsfs}%
\usepackage[title]{appendix}%
\usepackage{xcolor}%
\usepackage{textcomp}%
\usepackage{manyfoot}%
\usepackage{booktabs}%
\usepackage{algorithm}%
\usepackage{algorithmicx}%
\usepackage{algpseudocode}%
\usepackage{listings}%
\usepackage{xcolor}
\usepackage{soul}
\usepackage{pdfpages}
\usepackage{soul}
\soulregister\ref7
\soulregister\cite7

\theoremstyle{thmstyleone}%
\theoremstyle{thmstyletwo}%

\theoremstyle{thmstylethree}%

\begin{document}

\title[Article Title]{Optical fiber shape sensing using picosecond spatiotemporal single-photon reflectometry}
\author[1,*]{Harikumar K Chandrasekharan}
\author[1]{Saba N. Khan}
\author[1]{Muhammad Afnan Ansari}
\author[1]{Robert R. Thomson}
\affil[1]{Scottish Universities Physics Alliance, Institute of Photonics and Quantum Sciences, School of Engineering and Physical Sciences, Heriot-Watt University, David Brewster Building, Edinburgh EH14 4AS, Scotland, UK}

\affil[*]{Corresponding Author; email: hk47@hw.ac.uk}


\abstract{Inferring the three-dimensional configuration of a flexible optical fiber from a single accessible end remains a fundamental challenge in photonic sensing. Existing approaches typically require specialised multicore fibers, coherent interferometry, or distributed spectral encoding. Here, we demonstrate a fundamentally different sensing paradigm in which fiber geometry is reconstructed from the spatiotemporal dynamics of returning single photons. Using a laser-engineered multimode fiber interrogated by a time-resolved $32\times32$ single-photon avalanche diode array, we show that fiber deformation imprints unique signatures onto the distribution and arrival time of back-reflected photons. We map these high-dimensional photon-return states onto a low-dimensional geometric manifold, enabling continuous bend-angle estimation and reconstruction of previously unseen two- and three-dimensional shapes from measurements acquired entirely at the proximal fiber end. Measurements exhibit a median pixel-wise signal-to-noise ratio of $\sim26$\,dB and sub-1\% repeatability variation. Our results establish multimode single-photon dynamics as an information-rich channel for distributed geometric sensing, introducing a new framework for single-ended fiber shape reconstruction and minimally invasive photonic navigation.}




\maketitle

\section{Introduction}\label{sec1}

Optical fiber shape and curvature sensing underpin a wide range of applications spanning structural health monitoring, soft robotics, biomedical instrumentation, minimally invasive surgery, and robot-assisted navigation~\cite{Lopez-Higuera:11,s21051818,9981765,doi:10.1089,FLORIS2021106508,article139}. Existing fiber-based shape sensing approaches predominantly rely on fiber Bragg gratings~\cite{inproceedingsWU}, multicore fibers~\cite{Moore:12}, interferometric interrogation~\cite{s23073436}, or optical frequency-domain reflectometry~\cite{s18041072}, where geometric perturbations are inferred through spectral or phase-sensitive measurements. Although these techniques can provide high sensitivity, they often require complex interrogation architectures, coherent detection schemes, dense spectral multiplexing, or specialized fiber geometries, which can limit scalability, robustness, and system simplicity~\cite{FLORIS2021106508}.

Recent advances in computational photonics and high-dimensional optical sensing have highlighted the potential of extracting information from complex optical fields using multidimensional detector architectures. In particular, single-photon avalanche diode (SPAD) arrays combined with time-correlated single-photon counting (TCSPC)~\cite{Becker2005} provide access to spatially and temporally resolved photon measurements with picosecond-scale temporal precision. Such systems have enabled new capabilities in ultrafast imaging~\cite{Genevieve,Krstajic:15,10.1063,Chandrasekharan2026}, quantum sensing~\cite{Donaldson:21}, and biophotonics~\cite{Poland:15,Kufcsak2017}. Previously, multiplexed coupling of multimode optical fields into SPAD arrays using photonic lantern architectures demonstrated how spatially distributed single-photon timing channels can preserve and interrogate high-dimensional optical information with single-mode temporal performance~\cite{article77}.

\textcolor{black}{In parallel, embedding scattering and reflection structures within optical fibers have emerged as promising platforms for distributed sensing and reflectometric interrogation~\cite{Fu:23,CHEN2023109463,Wang:12,Yu:25}. \textcolor{black}{Recently, femtosecond laser-inscribed orthogonal eccentric scatterers have enabled three-dimensional fiber shape reconstruction through curvature-dependent backscattering intensities measured using optical time-domain reflectometry (OTDR)~\cite{Luo2026}.} However, most existing approaches primarily analyse one-dimensional temporal traces or integrated reflection amplitudes, thereby discarding the potentially rich spatial structure of the returned optical field including signatures of twist and deformation. In multimode fibers, geometric perturbations can modify modal coupling, relative phase evolution and scattering pathways, producing deformation-dependent spatial and temporal back-reflection signatures that are not fully captured by conventional single-detector OTDR measurements~\cite{s18041072,6637057,Peng:14}. Recent data-driven fiber-sensing approaches have further demonstrated that complex deformation-dependent optical signatures can be exploited for shape reconstruction~\cite{manavi2024deep}. This raises the possibility that the complex multimode field itself can act as an information-bearing representation of fiber geometry, provided that its spatial and temporal evolution can be resolved.}

\textcolor{black}{Here, we introduce a novel single-photon reflectometry framework for fiber-shape sensing that combines laser-engineered reflection sites within an optical fiber with a two-dimensional, time-resolved SPAD detector array. Light launched into the proximal end of the fiber is partially reflected by multiple engineered reflection sites and recorded by a $32\times32$ SPAD array, where each pixel acts as an independent photon-timing channel with $\sim55$ ps temporal resolution~\cite{Li:10,Poland:15,article77,10.1063}. Photon arrival time provides longitudinal discrimination along the engineered sensing region, while the SPAD array preserves the transverse spatial distribution of the returned light. Rather than reducing these measurements to a conventional one-dimensional reflectometric trace, the system therefore captures a high-dimensional spatiotemporal photon-return distribution, $H(x,y,t)$, that evolves with fiber geometry.}

\textcolor{black}{Using this approach, we demonstrate a  bend-angle prediction, followed by reconstruction of complex two-dimensional and three-dimensional fiber shapes. Principal component analysis (PCA)~\cite{9781118445112,RODRIGUEZGAR,Jolliffe2016} provides the central dimensionality-reduction framework, transforming the high-dimensional spatiotemporal photon-return distributions into a compact representation in which geometry-dependent states become structured and distinguishable. These latent features are subsequently mapped to physical fiber coordinates through partial least square (PLS) regression models, enabling reconstruction from independent measurements and prediction of previously unseen configurations excluded from calibration. The progression from one-dimensional deformation to complex two- and three-dimensional profiles demonstrates that the spatiotemporal multimode back-reflection contains sufficient information to recover fiber geometry beyond conventional bend or curvature sensing.}

\textcolor{black}{Our results establish geometry-dependent multimode propagation as an information-bearing resource for fiber shape sensing. By combining distributed optical reflection with spatially and temporally resolved single-photon detection, the approach transforms complex multimode back-reflections into an accessible representation of fiber geometry using interrogation entirely from the proximal end. This provides a route towards single-ended three-dimensional shape sensing in flexible optical fibers, with potential applications where direct observation or distal access is difficult.}

\begin{figure*}[t]
		\centering
		{\includegraphics[ width=1.0\linewidth]{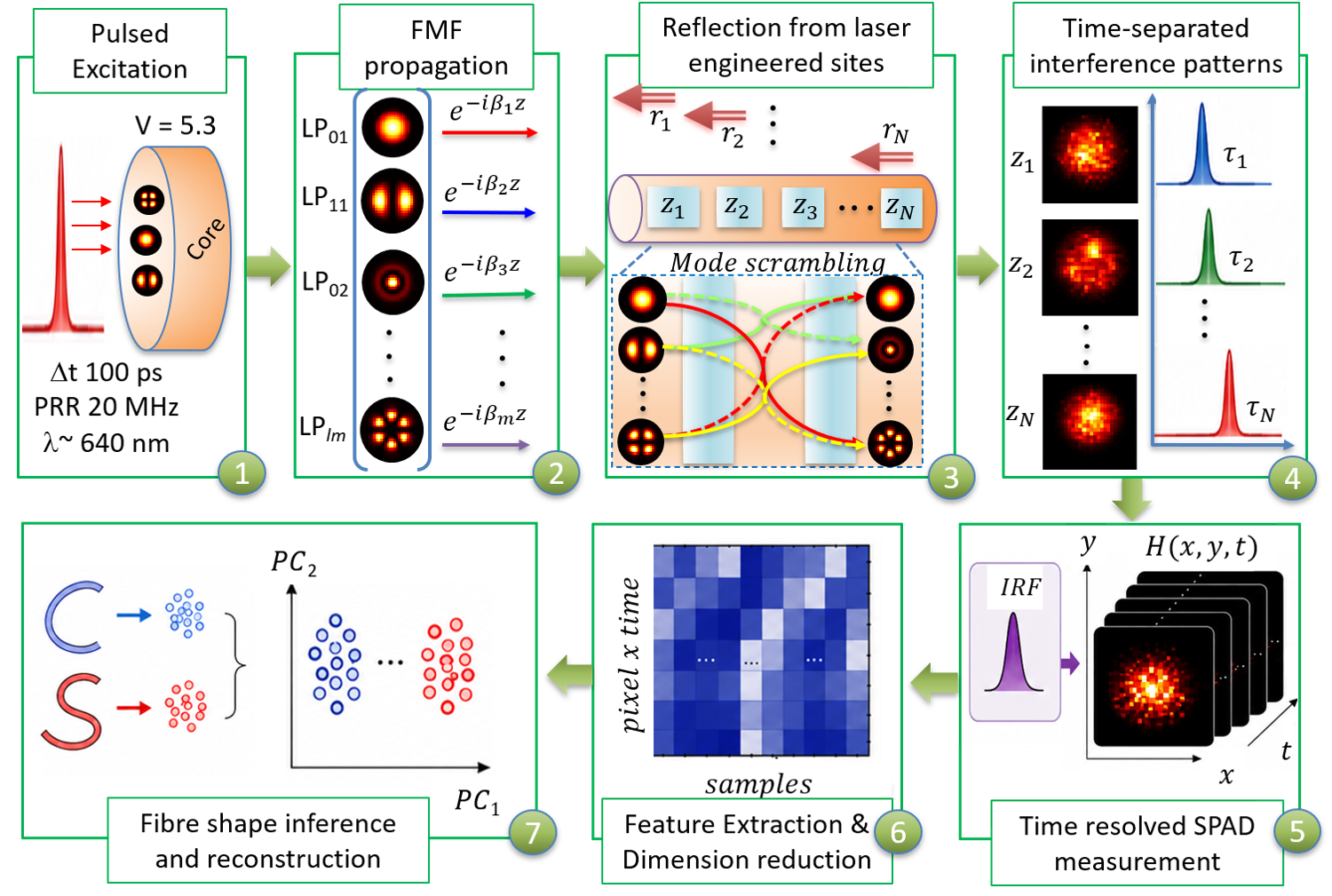}}
\caption{\textbf{Working principle of the proposed time-resolved SPAD-based fiber shape sensing.} A short optical pulse is launched into a few-mode fiber (FMF), exciting multiple guided modes that propagate along the fiber. Laser-inscribed reflection sites generate back-reflected signals, while fiber deformation induces intermodal coupling, thereby altering the modal interference pattern. Owing to the different axial positions of the reflection sites, the reflected photons arrive at distinct times, producing time-separated modal interference patterns. These spatio-temporal photon distributions are recorded using a time-resolved SPAD array to generate a three-dimensional dataset, $H(x,y,t)$. The time-gated SPAD distributions are subsequently processed for feature extraction and dimensionality reduction via principal component analysis (PCA), followed by partial least square (PLS) regression for fiber-shape reconstruction.}
\label{Fig:1}
\end{figure*} 
\section{Results}\label{sec:results}
\textbf{Working principle of spatiotemporal single-photon fiber shape sensing:}
\textcolor{black}{The operating principle of the proposed fiber shape-sensing framework is illustrated in Fig.\ref{Fig:1}. A short optical pulse is launched into a few-mode fiber (FMF), where it excites multiple guided spatial modes that propagate with mode-dependent phase evolution. As the pulse travels through the fiber, geometric perturbations modify the relative modal phases and induce mode coupling, thereby changing the spatial distribution of the propagating optical field. Spatially distributed femtosecond-laser-engineered reflection sites positioned at axial locations $z_1,z_2,\ldots,z_N$ locally sample these evolving multimode states and return a fraction of the optical field towards the proximal end. Since the reflection sites occupy different positions along the fiber, their photon returns occur at different temporal delays, allowing the longitudinal evolution of the multimode field to be interrogated through time-resolved detection.}

\textcolor{black}{The returned photons are recorded using a two-dimensional time-resolved SPAD array, producing a three-dimensional photon-count distribution $H(x,y,t)$, where the detector coordinates $(x,y)$ retain the spatial structure of the returning field and the photon arrival time $t$ provides depth information along the engineered sensing region. Fiber deformation is therefore encoded not simply through changes in total reflected intensity, but through the geometry-dependent evolution of spatial photon distributions across successive temporal positions. Following correction for the detector instrument response, these spatiotemporal photon-return topographies are converted into a high-dimensional measurement representation for feature extraction and dimensionality reduction. PCA is used to identify latent coordinates associated with fiber deformation, which, following calibration, are mapped to the corresponding fiber geometry. This enables reconstruction of increasingly complex two- and three-dimensional fiber shapes.}\\

\begin{figure*}[b]
		\centering
		{\includegraphics[ width=\linewidth]{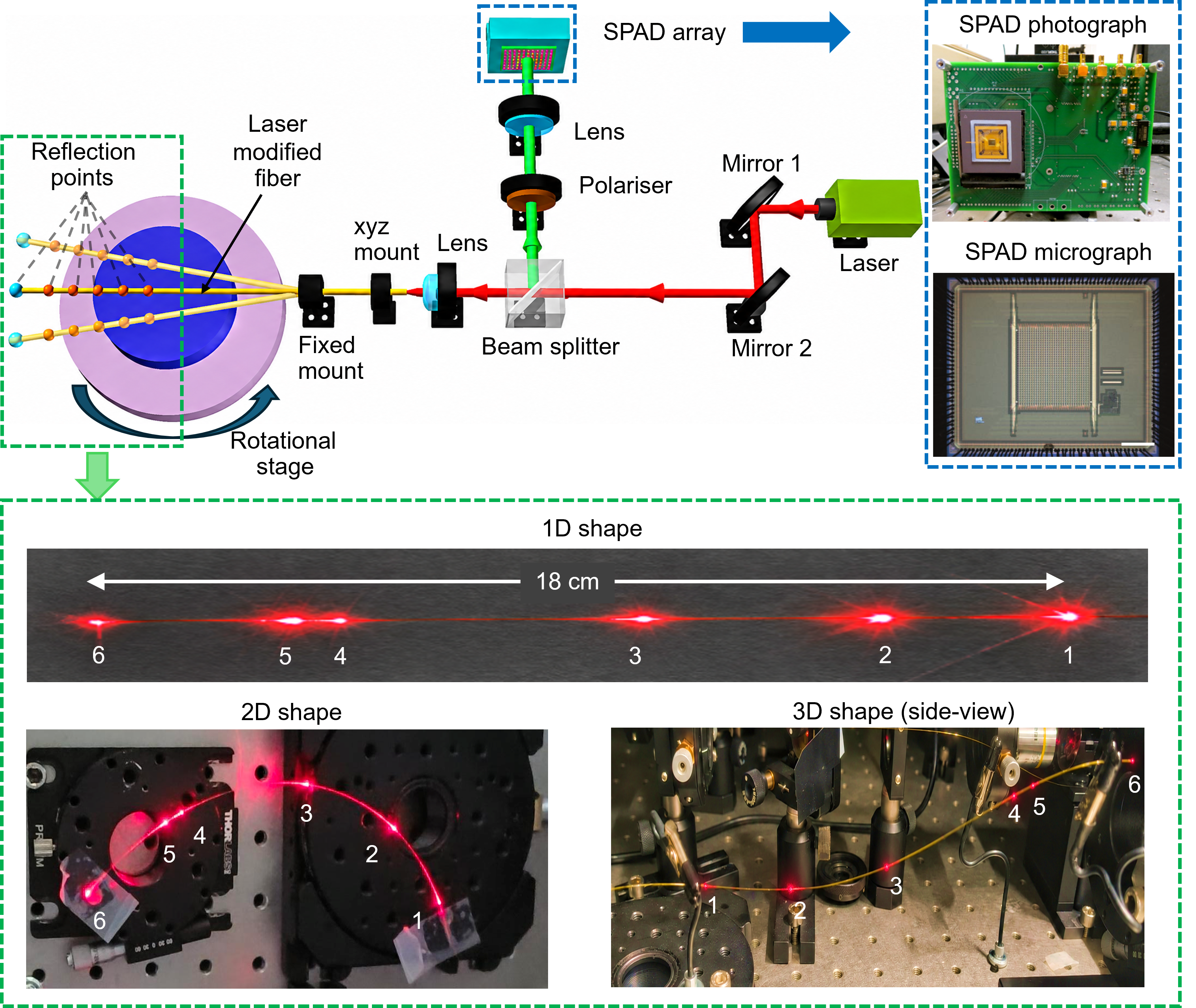}}
\caption{\textbf{Experimental setup used for spatiotemporal shape sensing with the SPAD array.} Pulsed light at 640\,nm coupled to a laser-modified SM1550P fiber through the transmission arm of a 50:50 non-polarising beamsplitter (BS). The fiber is secured using fixed mounts and mounted onto a rotational stage prior to the laser-engineered sensing region containing five reflection sites and a distal Fresnel reflector (site 6). Back-reflected photon signals (green beam path) are directed through the reflection arm of the BS onto the SPAD array. The green-outlined box shows an image of the laser-engineered fiber sensing region, together with photographs of representative 1D, 2D and 3D fiber shapes. The blue-outlined box shows (top) a photograph of the time-resolved 32×32 SPAD-array detection system operating in TCSPC mode and (bottom) a micrograph of the SPAD array’s sensing region. Scale bar, 800\,$\mu$m.}
\label{Fig:2}
\end{figure*}

\noindent\textbf{Experimental set-up:}
The experimental configuration used for spatiotemporal single-photon reflectometry in laser-engineered fibers is shown schematically in Fig.~\ref{Fig:2}. A pulsed laser source (PicoQuant laser) operating at a central wavelength of 640\,nm was used for time-resolved interrogation of engineered reflection sites within the fiber. The laser operated at a pulse repetition rate (PRR) of 20 MHz, producing pulses with a full width at half maximum (FWHM) duration of approximately 100\,ps and a spectral bandwidth of $\pm 5$\,nm. 

The optical signal was coupled into a laser-engineered SM1550P fiber placed in the transmission arm of a 50:50 non-polarising beamsplitter (BS). At 640\,nm, SM1550P fiber operates well below the wavelength range for single mode operation and has an estimated normalised frequency of $V\sim 5.30$ (core-diameter $\sim$ 9.0\,$\mu$m; NA $\sim$ 0.12). Under the weakly guiding approximation~\cite{Gloge,Snyder:78}, the fiber supports eight linearly polarised (LP) modes, namely LP$_{01}$, LP$_{11a}$, LP$_{11b}$, LP$_{21a}$, LP$_{21b}$, LP$_{02}$, LP$_{31a}$, and LP$_{31b}$. The fiber was approximately 1\,m long, with a \(\sim18\) cm sensing region containing five laser-engineered reflection sites and an additional Fresnel reflection site at the distal end facet, all used for shape reconstruction.

The back-reflected optical signal from the laser-engineered fiber was collected through the reflection arm of the beamsplitter and directed onto a time-resolved $32\times32$ SPAD array  operating in TCSPC mode (see Fig.~\ref{Fig:2}). A vertically oriented polariser was placed in the detection arm before the SPAD array to suppress unwanted polarisation-dependent background contributions and improve stability of the measured photon-return distributions. The SPAD array consists of pixels arranged on a square grid with a 50\,$\mu$m pixel pitch and an active photosensitive diameter of approximately 7\,$\mu$m per pixel, operating over the wavelength range 350--1050\,nm. 

Each SPAD pixel contains a dedicated time-to-digital converter (TDC) for independent TCSPC measurements with a dynamic range of approximately 56\,ns and a timing bin duration of 55\,ps~\cite{6176268,Krstajic:15}. Due to variability among the individual TDCs across the SPAD array, the pixel-wise instrument response function (IRF) FWHM ranged from approximately 381\,ps to 530\,ps (see Supplementary Note 1), enabling picosecond-scale temporal discrimination of the back-reflected photon-return signatures.\\

\noindent\textbf{Spatiotemporal photon-return dynamics in laser-engineered fibers:} Unlike conventional single-channel OTDR, where the detected signal is reduced to a one-dimensional temporal trace, the present system records a multidimensional spatiotemporal photon distribution $H(x,y,t)$ across the detector array. Consequently, geometric perturbations of the fiber produce not only temporal changes in the reflected signal amplitudes, but also systematic evolution of the spatial photon-return profiles measured across the SPAD pixels.
\begin{figure*}[b]
		\centering
		{\includegraphics[ width=\linewidth]{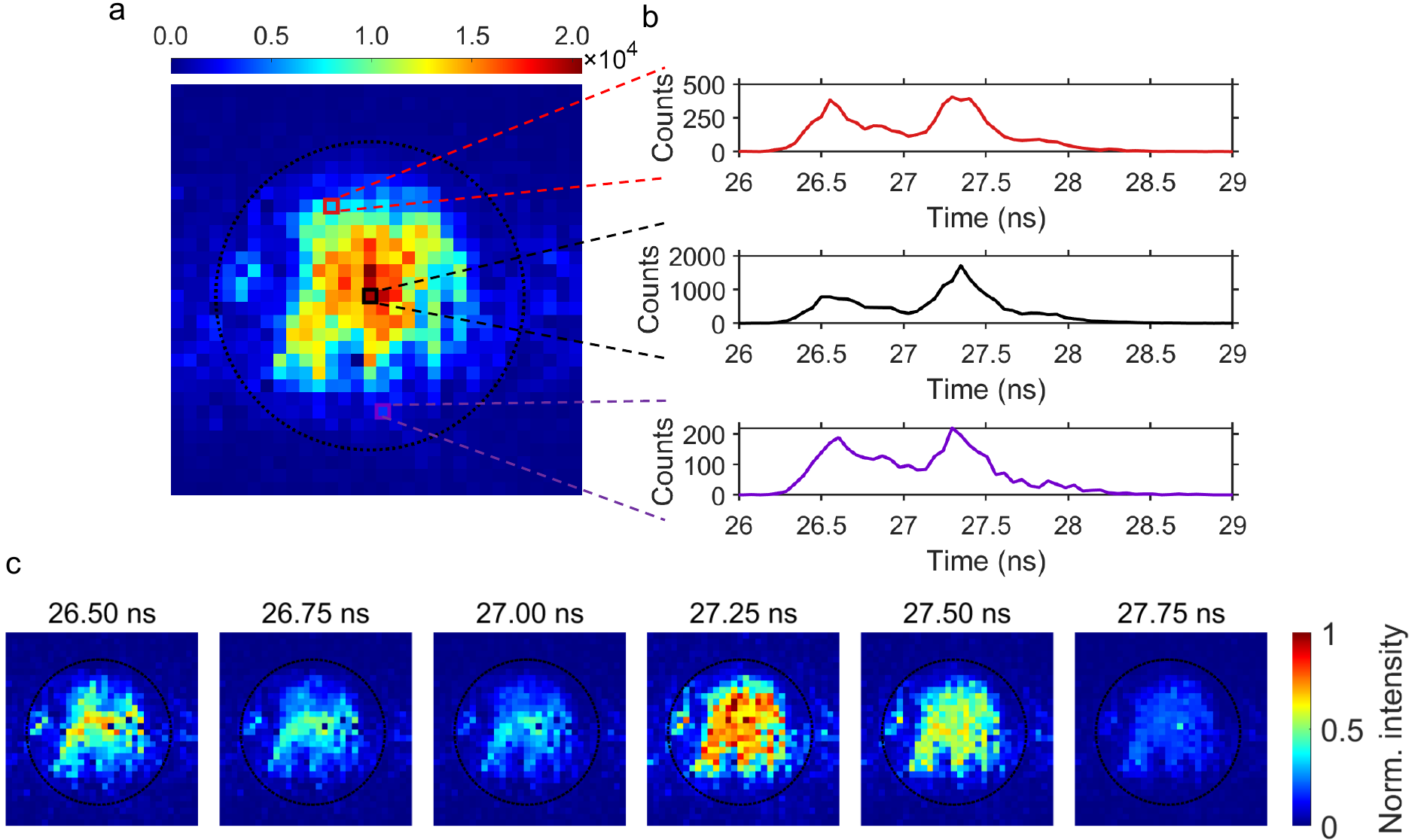}}
\caption{\textbf{Spatiotemporal photon-return dynamics within the laser-engineered sensing region.} \textbf{a} Summed photon-count distribution integrated over the temporally gated sensing window (26--28\,ns), showing the spatial distribution of back-reflected photons across the SPAD array. Square boxes in \textbf{a} indicate the representative SPAD pixels selected for temporal analysis. The dotted circle denotes the imaging region of interest used for subsequent analysis, consisting of 441 pixels. \textbf{b} TCSPC histograms from three representative SPAD pixels located at different positions across the sensing region, exhibiting distinct temporal photon-return profiles and spatially heterogeneous multimode photon dynamics. \textbf{c} Spatially resolved SPAD photon-return profiles extracted at successive 250\,ps temporal intervals across the sensing region. The dotted circle in the images denotes the approximate sensing region.}
\label{Fig:3}
\end{figure*}  
As a first demonstration, we investigated the intrinsic spatiotemporal photon-return dynamics generated by the laser-engineered reflection sites within the fiber. To introduce controlled photon redistribution through mode scrambling, the fiber sensing region was mounted in a C-shaped configuration. The bending-induced perturbations modify modal propagation, altering the spatial and temporal structure of the back-reflected photon distributions recorded by the SPAD array. In these measurements, we acquired 20,160 frames over 15 s, although we later demonstrate that the acquisition time can be reduced to 0.34 s while retaining a median signal-to-noise ratio (SNR) of up to 17 dB.

\textcolor{black}{In the reflection-mode imaging geometry, light reflected from the proximal fiber facet produces a strong, temporally distinct peak across the SPAD array. Although this peak is well separated from the sensing region ($\approx$ 70\,cm further along the fiber), its high photon count can drive the detector beyond the single-photon operating regime. The proximal reflection was therefore suppressed using index-matching gel and excluded by temporal gating. The isolated photon-return signal from the laser-engineered region of the fiber extends over $\approx$ 2\,ns, corresponding to an effective sensing length of $\sim18$\,cm.}

Figure~\ref{Fig:3}a shows the spatial distribution of back-reflected photon counts across the SPAD array, integrated over the 26–28 ns sensing window. Square boxes mark three representative pixels whose TCSPC histograms exhibit distinct temporal photon-return profiles (Fig. ~\ref{Fig:3}b), providing spatially resolved temporal information for inferring fiber deformation. The multidimensional SPAD measurements further enabled spatial photon-return profiles to be resolved at temporal intervals approaching the detector timing resolution of 55\,ps. In the present measurements, spatially resolved SPAD profiles were extracted at successive 250\,ps intervals across the engineered reflection region (as shown in Fig.~\ref{Fig:3}c), although finer temporal slicing down to the 55\,ps timing resolution is achievable in principle~\cite{Chandrasekharan2026}. The measured photon-return geometry evolves systematically across the reflection trace, corresponding to spatial propagation intervals of only a few centimeters along the fiber sensing region.\\

\noindent\textbf{Spatiotemporal photon-return dynamics under controlled fiber deformation:} 
To investigate the sensitivity of the multidimensional photon-return distributions to localised fiber deformation, the laser-engineered sensing region was mounted in a controlled C-shaped configuration and subjected to progressive cantilever loading. Incremental cantilever displacement introduced increasing mechanical perturbation within the bent sensing region (between sites 2 and 3), thereby modifying the underlying multimode propagation and scattering dynamics within the fiber. For each deformation state, time-resolved SPAD measurements were acquired across the full sensing region and temporally gated to isolate the engineered photon-return signatures from the proximal facet reflection.

\begin{figure*}[t]
		\centering
		{\includegraphics[ width=\linewidth]{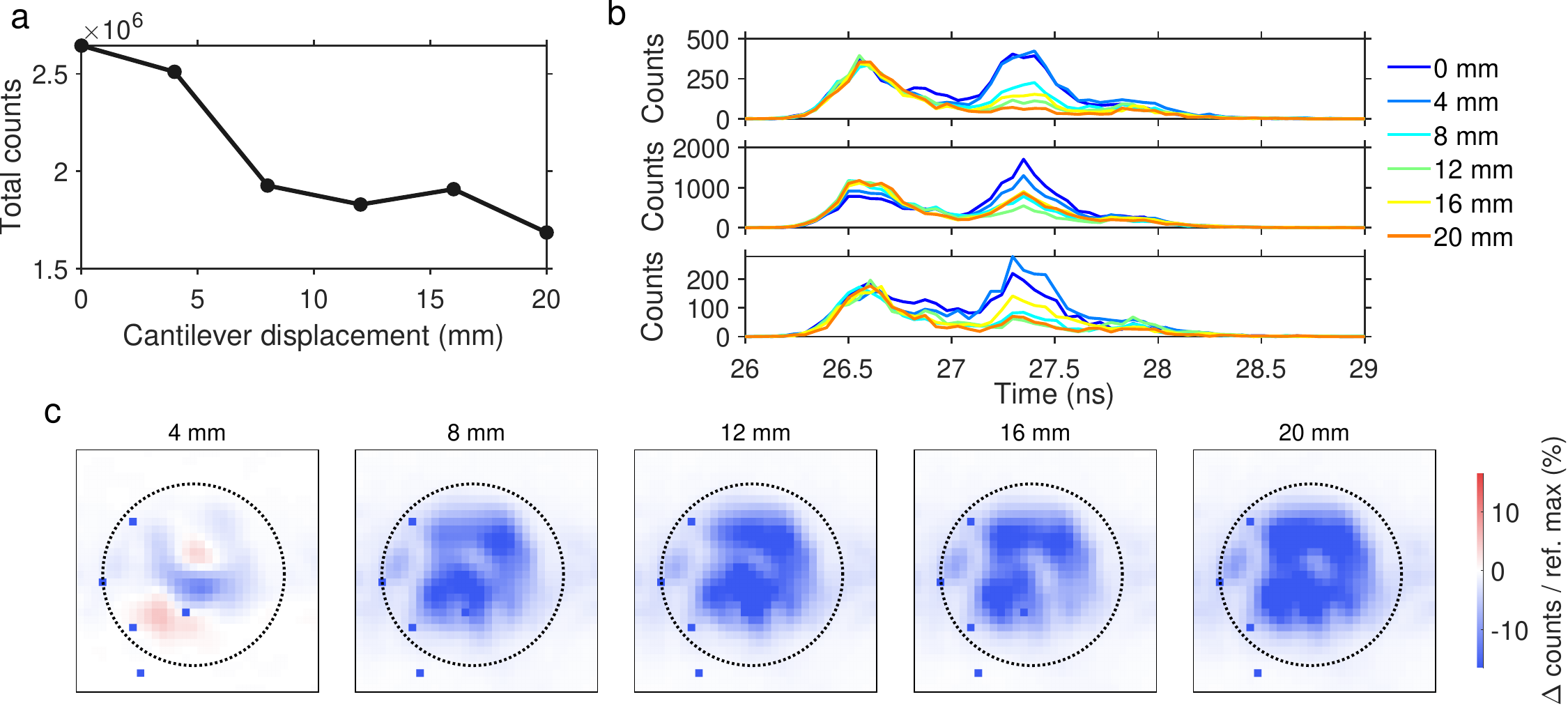}}
\caption{\textbf{Deformation-induced evolution of multidimensional photon-return distributions under controlled cantilever loading.} \textbf{a} Time-gated photon counts integrated over the laser-engineered sensing region as a function of cantilever displacement, showing modulation of the photon-return signal under increasing fiber deformation. \textbf{b} Representative TCSPC traces from three SPAD pixels within the sensing region (same pixels from Fig.~\ref{Fig:3}), demonstrating deformation-dependent reshaping of local temporal photon-return dynamics. \textbf{c} Spatial redistribution maps at the SPAD array for increasing cantilever displacement relative to the undeformed reference state (0\,mm). The maps were calculated as normalised differential photon-count distributions integrated over the 26--28\,ns sensing window, followed by Gaussian spatial smoothing to visualise deformation-induced redistribution of photon-return states across the SPAD array.}
\label{Fig:4}
\end{figure*}

Figure~\ref{Fig:4}a shows the total gated photon counts integrated across the sensing region as a function of cantilever displacement. Increasing deformation produced a systematic reduction in the overall photon-return intensity, indicating that local fiber perturbations strongly influence the photon dynamics. Beyond global intensity modulation, the deformation also induced substantial reshaping of the local temporal photon-return signatures recorded across the SPAD array. Representative temporal traces from three SPAD pixels are shown in Fig.~\ref{Fig:4}b, where distinct deformation-dependent evolution of the photon-return profiles can be observed. Importantly, different SPAD pixels exhibited markedly different temporal responses under identical mechanical perturbations, highlighting the inherently multidimensional and spatially heterogeneous nature of the photon redistribution.

\begin{figure*}[b]
		\centering
		{\includegraphics[ width=0.9\linewidth]{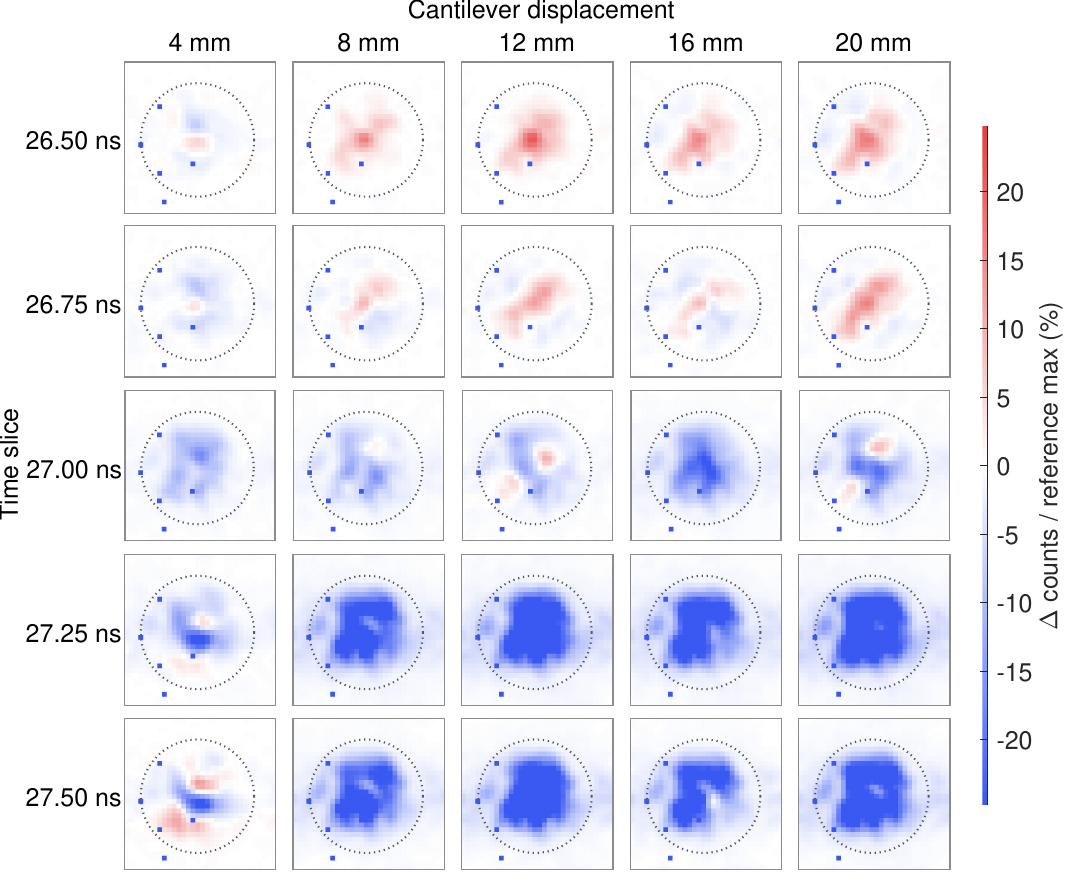}}
\caption{\textbf{Time-resolved deformation-induced normalised photon redistribution within the laser-engineered sensing region.}  Spatial redistribution maps at the
SPAD array acquired at successive temporal positions across the engineered reflection trace for increasing cantilever displacements. Each row corresponds to a fixed temporal slice of the photon-return signal, while each column represents a different cantilever displacement relative to the undeformed reference state (0 mm). Maps show the percentage change in photon counts relative to the corresponding temporal slice of the reference measurement. Distinct redistribution patterns emerge across both space and time, demonstrating that fiber deformation perturbs the multidimensional photon-return dynamics in a temporally dependent manner. The dotted circle indicates the region of interest used for subsequent analysis.}
\label{Fig:5}
\end{figure*}

The corresponding spatial redistribution maps are presented in Fig.~\ref{Fig:4}c, showing normalised differential photon-count distributions relative to the undeformed fiber state. Progressive cantilever loading generated highly structured spatial redistribution patterns across the SPAD array, revealing that fiber deformation does not simply attenuate the reflected signal uniformly, but instead induces correlated evolution of complex multimode photon-return states. These results demonstrate that localised fiber deformation can be encoded through coupled spatial and temporal photon-return dynamics, providing a high-dimensional optical measurement space suitable for geometric shape reconstruction.

While the integrated photon-return measurements in Fig.~\ref{Fig:4} demonstrate that fiber deformation produces measurable changes in both the spatial photon-count distribution and temporal response, the underlying dynamics become significantly more apparent when the photon-return states are resolved at individual temporal positions across the reflection sites. Figure~\ref{Fig:5} presents deformation-induced normalised photon redistribution maps extracted at successive temporal slices within the laser-engineered sensing region. Distinct redistribution patterns emerge as a function of both time and cantilever displacement, revealing that the applied mechanical stress does not simply attenuate the reflected signal, but instead perturbs the underlying multimode impulse response of the fiber. These changes arise from a combination of deformation-induced scattering losses, modal coupling, and multimode interference effects, resulting in temporally dependent photon redistribution across the SPAD array. Positive and negative redistribution regions appear simultaneously across the SPAD array, indicating deformation-driven reallocation of optical power between competing propagation pathways and modal states.

Notably, the strongest perturbations occur in the later portions of the reflection profile, where increasing cantilever displacement produces large-scale redistribution of photon-return states throughout the sensing region. Since the cantilever load was applied between reflection points 2 and 3, these observations suggest that local mechanical deformation preferentially modifies downstream propagation dynamics rather than generating a uniform response across all reflection sites. The measurements therefore provide direct experimental evidence that fiber deformation induces temporally dependent mode scrambling and photon redistribution within the laser-engineered structure. More broadly, Fig.~\ref{Fig:5} highlights a central capability of the proposed framework: the ability to directly visualise latent spatiotemporal photon dynamics that remain inaccessible to conventional integrated reflectometric measurements due to intensity averaging. These multidimensional photon-return states form the physical basis for the fiber shape reconstruction approaches presented in the following sections.
\\

\noindent\textbf{Spatiotemporal photon-return distributions in a one-dimensional shape sensing:} 
While the previous section established that the laser-engineered fiber generates structured multidimensional photon-return dynamics across the SPAD array, the central capability of the proposed framework lies in reconstructing latent geometric states directly from these high-dimensional single-photon measurements. To investigate geometric sensitivity and shape-retrieval performance, the laser-engineered sensing region was mounted in a straight configuration on a rotational stage, while the proximal fiber section was secured using a fixed mount positioned before the rotation axis (see Fig.\ref  {Fig:1}). Controlled fiber deformations were subsequently introduced through systematic clockwise and anticlockwise angular rotations of the sensing section, enabling repeatability and hysteresis of the multidimensional photon-return states to be evaluated. For each angular configuration, time-resolved SPAD frames corresponding to the sensing region and distal Fresnel reflection were recorded and analysed.

\begin{figure*}[t]
		\centering
		{\includegraphics[width=\linewidth]{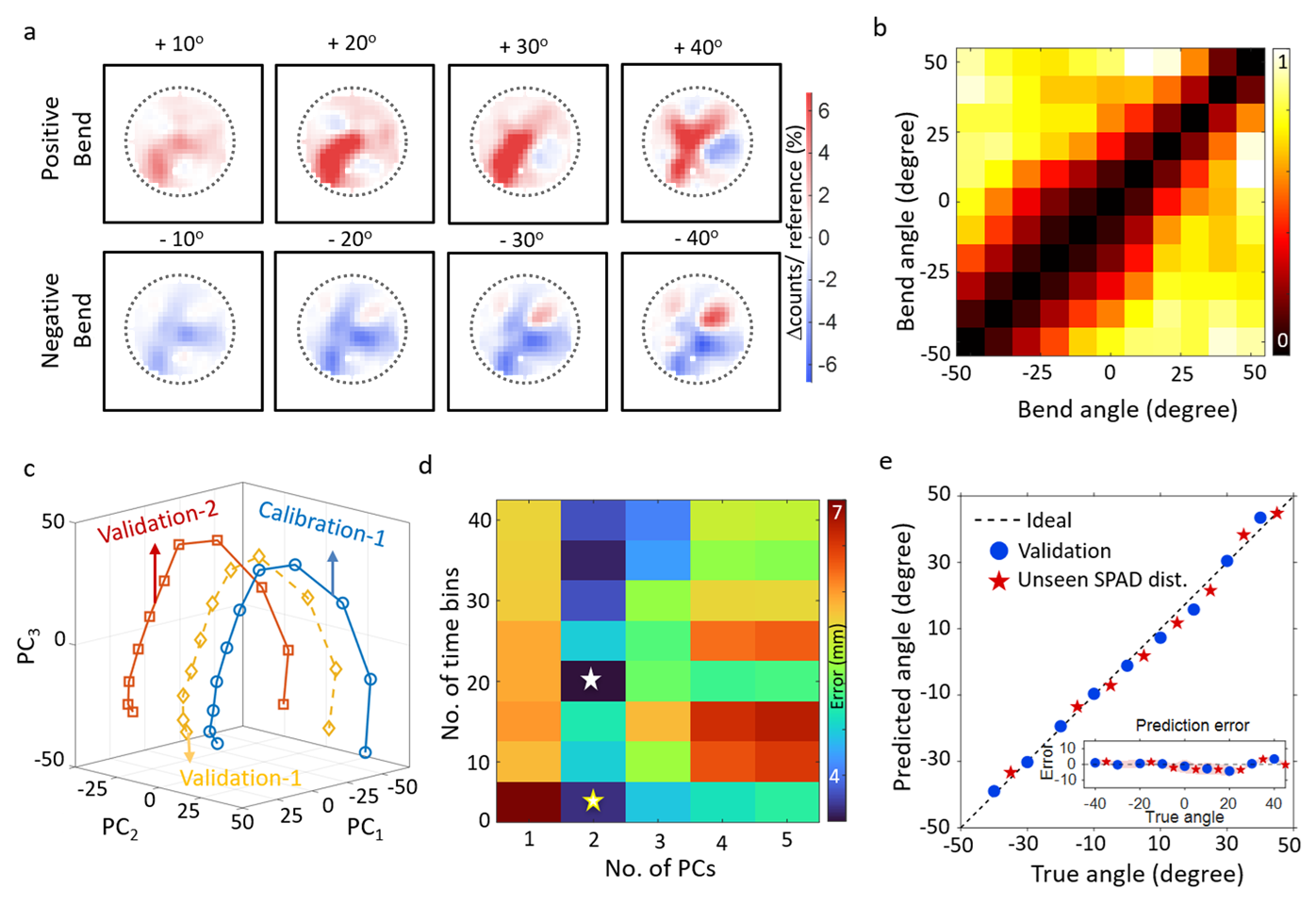}}
		\caption{\textbf{Time-gated SPAD distribution analysis and bend-angle prediction.} \textbf{a} Representative changes in SPAD distributions relative to the fixed reference state for positive and negative bend angles. Red and blue indicate an increase and decrease in photon counts, respectively. \textbf{b} normalised dissimilarity between SPAD patterns at different bend angles. \textbf{c} Principal-component trajectories of the calibration and independent validation datasets. \textbf{d} Minimum mean absolute error (MAE) obtained by optimising the number of time bins, PCA components, and regression model. \textbf{e} Predicted versus true bend angles for independent validation measurements and intermediate angles excluded from calibration (stars). The dashed line represents ideal prediction; the inset shows the prediction error, with shading indicating variation among repeated measurements.}
	\label{fig:6}
	\end{figure*}
    
The acquired spatiotemporal photon-return distributions were subsequently analysed using PCA to investigate the latent fiber bend angle. Unlike conventional reflectometric techniques that compress the returning signal into a single integrated temporal trace, the present framework preserves the spatially resolved temporal dynamics measured across the SPAD array. Consequently, one dimensional fiber bending produces correlated changes in high-dimensional photon-return patterns that can be represented in a reduced-dimensional space. 
Figure ~\ref{fig:6} demonstrates the use of time-gated SPAD measurements for continuous bend-angle estimation, with opposite bending directions generating distinct responses (Fig.~\ref{fig:6}a). The dissimilarity matrix shows that differences between the SPAD patterns generally increased with angular separation (Fig.~\ref{fig:6}b). Consistent with this behaviour, PCA revealed smooth and ordered trajectories with bend angle, while independent measurements performed on different days closely followed the corresponding calibration trajectories (Fig.~\ref{fig:6}c). Optimisation of the temporal and PCA parameters identified the feature configuration that minimised the prediction error (Fig.~\ref{fig:6}d). The resulting model showed close agreement between predicted and applied bend angles for both independent test measurements and intermediate angles excluded from calibration (Fig.\ref{fig:6}e), demonstrating continuous angle prediction and generalisation to previously unseen bending states. These results confirm that the spatiotemporal photon-return distributions encode robust and transferable geometric information, enabling accurate prediction of fiber deformation states directly from single-photon measurements. Having established that the time-gated SPAD distributions encode both the magnitude and direction of fiber bending, we next investigated whether these measurements could recover more complex two-dimensional fiber profiles. \\

\noindent\textbf{Spatiotemporal photon-return distributions for in-plane (two-dimensional) shape sensing:}
To investigate two-dimensional geometric reconstruction, the laser-engineered fiber was mounted between two independently controlled rotational stages positioned before the first engineered reflection site and at the distal fiber end, respectively. By changing the relative angular displacement between the two stages, the fiber could be continuously transformed into controlled C-shaped conformations with varying radius of curvatures. For each geometric configuration, multidimensional SPAD photon-return distributions were recorded and independently repeated to evaluate reconstruction precision and repeatability.
\begin{figure*}[t!]
		\centering		{\includegraphics[width=1.0\linewidth]{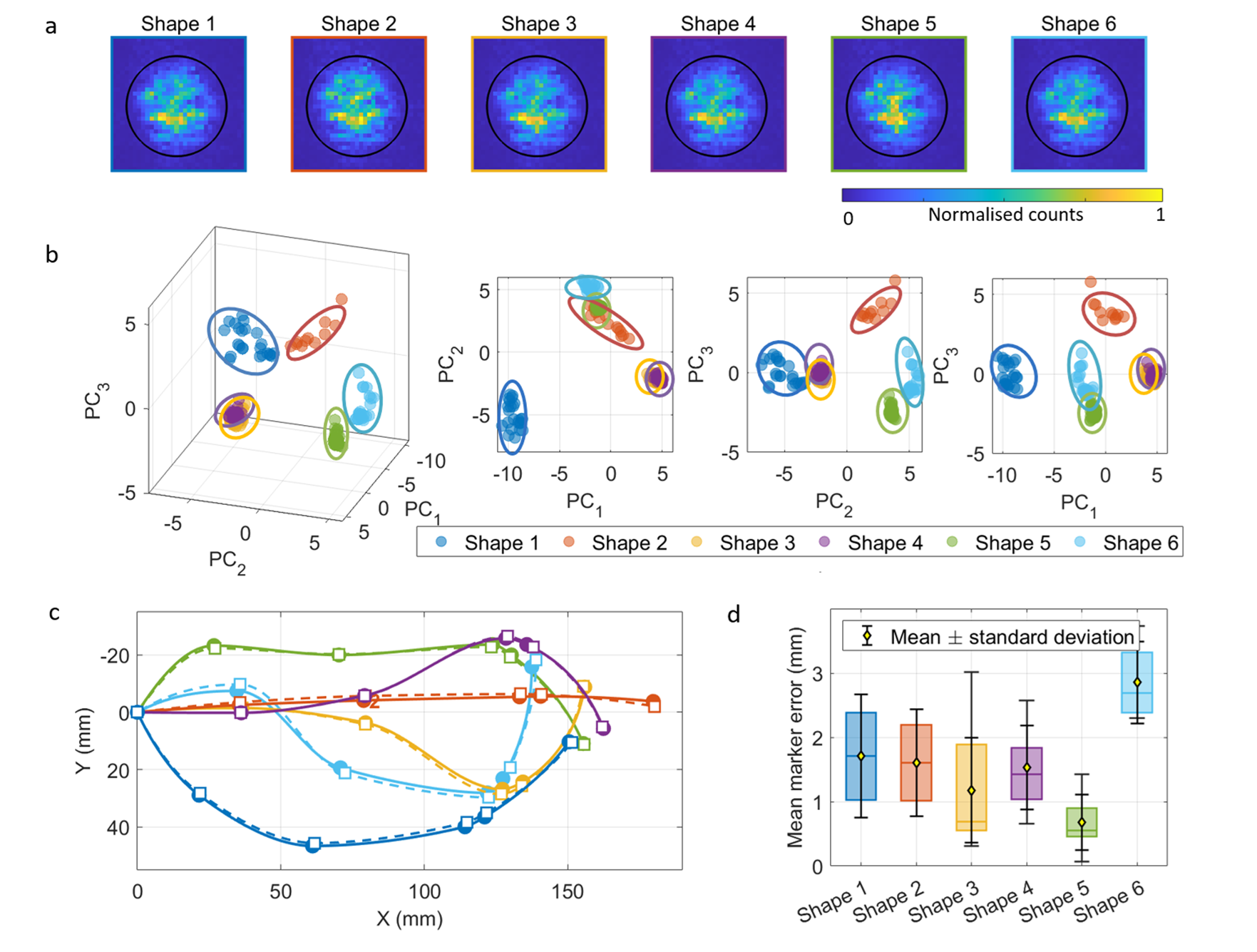}}
	\caption{\textbf{Spatiotemporal reconstruction of two-dimensional fiber shapes.} \textbf{a} Representative SPAD photon-count distributions measured for the six different fiber-shapes. The circles indicate the spatial regions of interest used in the analysis. \textbf{b} Distribution of the time-gated SPAD measurements in the three-dimensional principal-component space, together with the corresponding two dimensional principal component projections. Colors identify the six different shapes. \textbf{c} Comparison of the ground truth fiber profiles (circles and solid curves) with the profiles reconstructed from the SPAD measurements (squares and dashed curves) using the PCA-PLS regression models. \textbf{d} Distribution of the mean reconstruction error for each shape, calculated from the Euclidean distances between the measured and predicted marker positions. Yellow diamonds and error bars represent the mean and standard deviation, respectively.}
	\label{fig:7}
	\end{figure*}
\textcolor{black}{The acquired spatiotemporal photon-return measurements were subsequently analysed using PCA to identify the dominant latent geometric shape. The first few principal components captured the majority of the variance within the measured photon-return manifold, indicating that the complex multimode scattering dynamics evolve within a relatively low-dimensional latent space despite the high-dimensional SPAD measurement basis.}

\textcolor{black}{The two-dimensional shape-sensing results are summarised in Fig. ~\ref{fig:7}. Representative raw, time-gated SPAD photon-count distributions for the six fiber-shape types are shown in Fig. ~\ref{fig:7}a. Although the SPAD-intensity distributions appear broadly similar, small shape-dependent changes in the photon distribution provide sufficient information for classification and reconstruction. PCA of the gated spatiotemporal data reveals distinct clustering of the six shape types in the first three principal components and their corresponding two-dimensional projections (Fig. ~\ref{fig:7}b). While the projected points for shape types 3 and 4 lie close together in PCA space, the two remain clearly distinguishable (see Supplementary Note 2). Their proximity arises from their similar fiber geometries across the first three reflection sites (see fig. ~\ref{fig:7}c ). The ground-truth marker positions were obtained from the bright scattering points corresponding to the laser-modified sites of the fiber. Figure ~\ref{fig:7}c compares the camera-measured ground-truth fiber profiles with those reconstructed from the SPAD measurements using the type-specific PCA–PLS (partial least square) models. The reconstructed profiles closely follow the measured marker positions and reproduce the characteristic geometries of the six different shape types. Multiple fiber profiles with varying curvature along their length were tested within each shape type. For visual clarity, Fig.~\ref{fig:7}c shows only six representative shapes. Fig.~\ref{fig:7}d shows the distribution of mean marker errors across all tests for each shape type, with yellow diamonds and error bars indicating the mean and standard deviation, respectively.  Across the six types, the mean errors remain below approximately 3 mm, with shape type 5 exhibiting the lowest error and shape type 6 showing the largest error and variability. These results demonstrate that subtle variations in the time-gated SPAD response can be used to distinguish and reconstruct a range of two-dimensional fiber configurations.}\\


    \begin{figure*}[t!]
		\centering		{\includegraphics[width=1.0\linewidth]{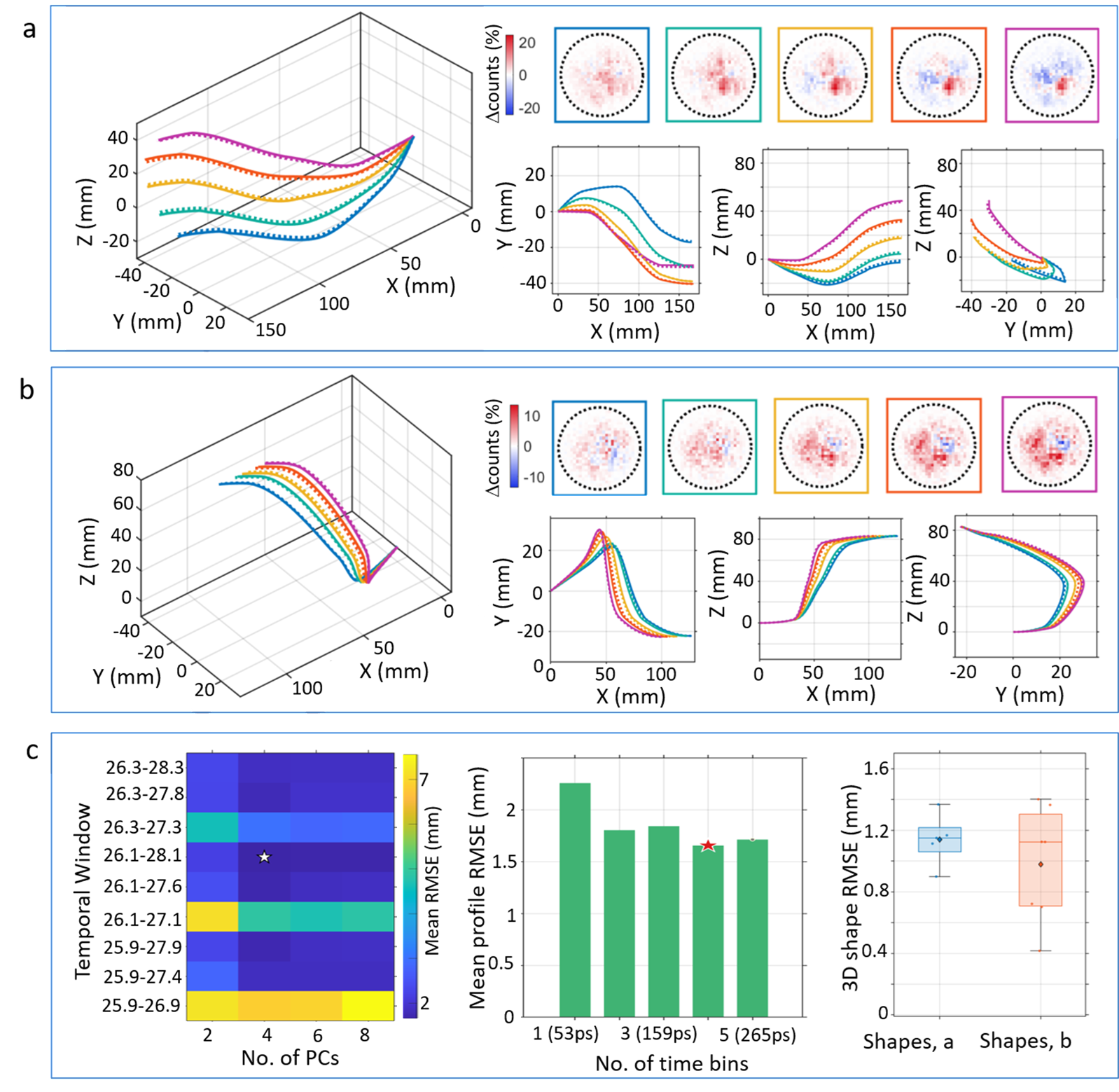}}
	\caption{\textbf{Spatiotemporal reconstruction of three-dimensional fiber shapes.} \textbf{a,b} Reconstruction results for two representative shape types. For each type, the ground truth shapes (solid curves) and corresponding reconstructions (dotted curves) are shown in three dimensions and as projections onto the three orthogonal planes. Colours identify the five shape configurations and correspond to their respective spatially resolved SPAD profiles shown above the projections. \textbf{c} Parameter optimisation and quantitative evaluation of the reconstructed shapes, showing the parameter-search error map, selected number of principal components, and reconstruction-error distributions for the two shape types. The star symbols indicate the selected model parameters.}
	\label{fig:8}
	\end{figure*}    
\noindent\textbf{Spatiotemporal photon-return distributions for three-dimensional shape sensing:}
To advance towards practical shape sensing applications, we next extended the time-gated SPAD framework from in-plane shape sensing to three-dimensional reconstruction of more complex fiber geometries involving twist. Two representative shape types were investigated independently. For each shape type, five configurations spanning a continuous progression of the fiber geometry were reconstructed from their time-resolved SPAD measurements. The relative SPAD maps show that changes in fiber configuration produced distinct spatial distributions of the detected photon counts. Importantly, these changes were not limited to the total collected intensity but altered the spatial pattern across the detector, providing configuration-dependent information for shape recovery. 

Figures  ~\ref{fig:8}a \& b present the reconstruction of the two different shape types. The predicted three-dimensional profiles closely follow the corresponding camera-derived ground truth profiles across all five different (unseen) configurations. Agreement is also maintained in the three orthogonal projections, demonstrating that the model recovers changes in both transverse coordinates rather than only the dominant bending plane. In fiber-optic shape sensing systems, fixing the probe at both ends generally introduces unavoidable twist. In our example shape types, this twist was inherently present and varied as we gradually changed the shapes. Nevertheless, we successfully recovered the 3D shapes, including their varying twist, from the SPAD photon-count distributions for five different shapes within each shape type. 
In both panels (a and b), the colour coding links each reconstructed profile to its corresponding relative SPAD distribution. 
The temporal and PCA parameters were selected using training-only leave-one-out cross-validation (Fig.  ~\ref{fig:8}c). The optimization map shows the reconstruction error obtained for different temporal aggregation and PCA configurations, with the selected setting indicated by the star. Comparison across the investigated temporal aggregations identified the configuration that provided the lowest validation error without using the independent test data. The resulting models were subsequently evaluated on independent measurements. The error distributions show consistent reconstruction performance for both shape types, although the broader distribution for the second types indicates greater variability for the more strongly curved geometry. Together, these results demonstrate that the time-gated SPAD framework can recover continuous three-dimensional fiber profiles across distinct shape types and generalise to measurements excluded from model training.{\\}

\noindent\textbf{Signal-to-noise ratio and dependence on acquisition time.}
\textcolor{black}{The SNR of the photon-return measurements was evaluated from ten independent acquisitions under conditions representative of the shape-sensing experiments, using the signal and pre-signal temporal windows defined in Supplementary Note~4. Across the circular detector region used throughout the analysis, the pixel-wise SNR, evaluated from the signal and noise statistics across ten repeated acquisitions, ranged from approximately 7.5 to 36\,dB, with a median value of $\sim26$\,dB under conditions representative of the shape-sensing experiments.} 

\textcolor{black}{The dependence of SNR on acquisition time was further evaluated using measurements acquired under the same experimental settings while varying the number of accumulated frames from 480 to 20,160 (Supplementary Fig.~S7a). The array-integrated SNR increased with frame accumulation and exhibited a scaling of 5.70\,dB per tenfold increase in frame number ($R^2=0.986$), close to the 5\,dB increase expected from $\sqrt{N}$ scaling for statistically independent noise. Notably, at 480 frames---the minimum frame setting permitted by the current SPAD acquisition software---the acquisition time was approximately 0.34\,s, while retaining a clear spatially resolved photon-return signal with SNR ranging from 4.86 to 24.45~dB, with a median value of 17.38~dB. This demonstrates that the system can operate in the sub-second acquisition regime under the present experimental configuration. The spatial SNR distribution and its dependence on acquisition time are presented in Supplementary Fig.~S7b.}{\\}

\noindent\textbf{Measurement stability and repeatability.}
\textcolor{black}{Measurement stability and repeatability were evaluated over ten independent acquisitions. Analysis was restricted to a circular region of interest encompassing the photon-return distribution, with only pixels satisfying the applied SNR threshold included. For a fixed fiber configuration, SNR-qualified pixels yielded an array-integrated relative standard deviation (RSD) of 0.78\% and a median pixel-wise RSD of 2.69\%. Repeatability was independently evaluated by displacing the positioning stages and subsequently returning the fiber to the same reference configuration before each acquisition. Across ten repeated measurements, qualified pixels yielded an array-integrated RSD of 0.86\% and a median pixel-wise RSD of 2.80\%. The close agreement between these measurements demonstrates good stability and repeatability of the photon-return measurements (see Supplementary Note 5).}
 
\section{Discussion and conclusions}

 \textcolor{black}{Recovering the three-dimensional trajectory of a flexible fiber from a single accessible end is important where direct observation of the fiber is difficult or impossible. Our results establish complex multimode propagation as an information-bearing mechanism for fiber shape sensing, turning optical complexity that is conventionally suppressed into a resource for geometric reconstruction. Fiber deformation modifies modal coupling and interference, producing geometry-dependent photon distributions that evolve along the propagation path. Laser-engineered reflection sites sample these evolving optical states, while photon arrival time provides longitudinal discrimination and the SPAD array preserves their transverse spatial distribution.}

\textcolor{black}{Together, these measurements form a high-dimensional spatiotemporal photon-return field, $H(x,y,t)$, that encodes fiber geometry and can be interrogated entirely from the proximal end. Shape information is therefore recovered from photon-count distributions without phase-resolved coherent detection, dense spectral interrogation, or distributed strain measurements. By shifting the complexity of shape recovery from specialised fiber architectures and interrogation schemes to multidimensional detection and computational decoding, this approach provides a distinct route towards compact, single-ended fiber shape sensing.}

 \textcolor{black}{Extending bend angle prediction to 2D and 3D shape reconstruction demonstrates that the encoding extends beyond a single deformation parameter. In-plane profiles were reconstructed with millimetre-scale position errors, while previously unseen bending states and three-dimensional out-of-plane trajectories excluded from model training were successfully recovered. This generalisation beyond calibrated configurations demonstrates that the spatiotemporal photon-return field encodes sufficient geometric information for reconstructing complex fiber trajectories, advancing the approach from bend or curvature detection towards visualisation of fiber's geometric shape and trajectory.}

 \textcolor{black}{The practical potential of this approach is further supported by the measurement quality and robustness of the photon-return signatures. Under conditions representative of the shape-sensing experiments, the spatially resolved measurements exhibited a median pixel-wise SNR of $\sim26$\,dB, while useful photon-return distributions were retained at the minimum accessible acquisition time of 0.34\,s, demonstrating sub-second operation. Importantly, repeated measurements of a static fiber configuration showed an array-integrated RSD of only 0.78\%, increasing only marginally to 0.86\% after displacement and repositioning of the fiber mount. Together, the high SNR, sub-second acquisition and sub-1\% integrated variability demonstrate that the geometric information encoded in the multimode photon-return field is both experimentally accessible and highly reproducible, providing a robust basis for computational shape reconstruction.}

 \textcolor{black}{The proposed approach offers several practical advantages over conventional fiber-shape sensing architectures. By operating with a single-core fiber and exploiting time-gated, spatially resolved photon detection, it avoids the multicore fan-out and multichannel spectral demodulation typically associated with multicore fiber Bragg grating sensors, while enabling deformation-sensitive optical signatures to be interrogated at multiple longitudinal positions. As the sensing information is extracted from spatiotemporal photon-count distributions rather than wavelength shifts, the approach may offer reduced sensitivity to certain sources of spectral drift, although its response to temperature and axial strain remains to be established. The present implementation nevertheless has limitations: longitudinal discrimination is constrained by the instrument response, temporal sampling and photon-counting statistics, while the number and spacing of usable reflection sites depend on optical loss, return strength and temporal overlap. Furthermore, reconstruction requires representative calibration, and neighbouring time gates may contain correlated information owing to the finite instrument response. Future work will therefore assess robustness to environmental perturbations, fiber-to-fiber variability and long-term drift, alongside improvements in photon-collection efficiency and acquisition speed.}

\textcolor{black}{More broadly, the combination of laser-engineered reflection regions and time-resolved multipixel single-photon detection provides a route towards accurate path visualisation using optical fibers. \textcolor{black}{Increasing the density of engineered reflection sites and exploiting additional optical degrees of freedom, particularly polarisation, could further enrich the geometric information available for reconstruction.} Together with advanced computational reconstruction approaches~\cite{aisy.202100067,articleYujiao}, these developments could enable real-time shape feedback in minimally invasive instruments~\cite{articleIssatayeva}, soft robotics~\cite{doi:10.1089}, structural monitoring~\cite{Lopez-Higuera:11}, and autonomous robotic systems~\cite{bookLeal}. By harnessing geometry-dependent multimode propagation rather than suppressing it, the approach offers a pathway towards single-ended optical fibers that simultaneously guide light and sense its own three-dimensional shape.}

\section*{Methods}\label{sec4}

\subsection*{TCSPC experimental details}

Time-correlated single-photon counting measurements were performed using a reverse start--stop timing architecture. The laser synchronization output (NIM signal) was routed to a time-tagger module (ID900, ID Quantique), which provided timing control and generated the common reference signal distributed to the SPAD array for TCSPC acquisition. To minimize detector saturation and photon pile-up effects~\cite{Harris1979SinglePhotonDS}, the photon flux was adjusted such that the average detection probability remained below approximately 10\% per laser pulse.

For each detected photon, the TDC recorded the interval $\Delta t$ between the photon detection event and the subsequent synchronization pulse. Photon arrival times within the laser repetition period were reconstructed according to

\begin{equation}
t_{\mathrm{arr}} = T_{\mathrm{rep}} - \Delta t,
\end{equation}

where $T_{\mathrm{rep}}$ denotes the laser repetition period. Arrival times were wrapped to the interval $[0,T_{\mathrm{rep}}]$ and accumulated into pixel-resolved TCSPC histograms, enabling reconstruction of the spatiotemporal photon-return distributions from the laser-engineered fiber.

As part of the detector calibration procedure, the IRF was characterized for every SPAD pixel. The measured IRF incorporates temporal broadening arising from the excitation pulse width, detector timing jitter, and delays introduced by the timing electronics and signal distribution network. To quantify these effects, the array was uniformly illuminated using the 640\,nm pulsed laser source and the FWHM of the TCSPC response was extracted on a pixel-by-pixel basis. Across the array, the measured IRF FWHM ranged from 381~ps to 530~ps, with a mean value of 470~ps (See Supplementary Note 1).

Pixel-dependent timing offsets associated with variations in the underlying TDC channels resulted in non-uniform temporal shifts between detector elements. To establish a common temporal reference, the peak arrival time of each pixel was determined from the calibration measurements and converted to integer multiples of the 55\,ps timing bin. These offsets were subsequently applied to all acquired TCSPC datasets, enabling temporal alignment of the photon-return traces across the entire SPAD array.
\subsection* {Femtosecond laser inscription of the shape sensing probe}

The shape-sensing probe was realised using commercially available polyimide-coated SM1550P fiber (Thorlabs) by inscribing discrete high-refractive-index regions, referred to as reflection sites, within the fiber core through the intact coating. The refractive index modifications were inscribed using 1030-nm, 185-fs laser pulses with a pulse energy of 220 nJ at a repetition rate of 1 kHz. The laser beam was expanded and spatially shaped using a spatial light modulator, which also applied a phase correction to compensate for spherical aberration. The beam was then focused into the fiber core using a microscope objective with a numerical aperture of approximately 0.56. Half- and quarter-wave plates controlled the polarisation of the writing beam. A CCD camera positioned in the imaging path reflected by the dichroic mirror provided a side view of the fiber. As the translation stage was moved along the $z$-axis, an image-recognition algorithm analysed the intensity gradient along a vertical line profile in real time to locate the fiber core precisely.  
During laser inscription, the fiber was translated slowly relative to the focused laser beam using a computer-controlled translation stage. Each reflection site consisted of three to four parallel laser-modified regions inscribed within the fiber core. The individual regions were approximately 1.2 $\mu$m wide and 200 $\mu$m long. Multiple sets of these parallel modifications were inscribed along the length of the fiber to form a spatially distributed array of reflection sites. The resulting sites generated localised back-reflections while preserving the flexibility of the fiber. Their known spatial separation enabled the photon returns from different locations to be resolved temporally using the time-gated SPAD detection system.
\subsection*{SPAD data analysis and fiber-shape reconstruction}
\textcolor{black}{Time-gated SPAD measurements were analysed using PCA and PLS regression to classify and reconstruct the fiber shape. For each measurement, the background was subtracted, negative values were set to zero, and photon-arrival data within an optimised temporal window around 26–28 ns were retained. At the temporal resolution of 55 ps, each time bin contained a 32 × 32-pixel photon-count distribution. These distributions were normalised and vectorised while preserving their spatial and temporal information, and PCA was used for dimensionality reduction.}

\textcolor{black}{For two-dimensional sensing, a support vector machine with a radial-basis-function kernel classified each measurement into a shape type. A separate PCA–PLS regression model was trained for each type to map the reduced SPAD features onto camera-derived Cartesian coordinates of six markers positioned along the 180-mm fiber. For an unknown measurement, the predicted type selected the corresponding regression model, and the resulting shape scores were transformed back into marker coordinates.
For two and three-dimensional sensing, ground-truth marker coordinates were obtained by combining reflection sites (appearing as bright points) calibrated top- and side-camera projections. The two shape types were modelled independently using an adaptive ensemble of locally weighted and global ridge-regression models in PCA space. The local model was weighted more strongly for measurements close to the training distribution, whereas the global model provided a stable prediction for sparsely sampled or unfamiliar measurements. The temporal window, temporal aggregation, number of principal components, nearest-neighbour count, ridge regularisation and model parameters were selected exclusively from the training data using leave-one-out cross-validation, with the root-mean-square error (RMSE) as the optimisation criterion.
For both sensing configurations, continuous fiber profiles were generated from the predicted marker coordinates using shape-preserving piecewise cubic interpolation. Two-dimensional classification was assessed using five-fold cross-validation, while reconstruction was evaluated by leave-one-out cross-validation and independent test measurements. Accuracy was quantified using the Euclidean marker-position error; for 3D reconstruction, the full-profile RMSE and distal-tip error were also calculated.}

\subsection*{Numerical simulations: spatiotemporal backscattering in a laser-engineered fiber}

To investigate the physical origin of the deformation-dependent photon-return dynamics, numerical simulations were performed using full-vector coupled-mode theory (Supplementary Note 3). Complex transverse electric fields obtained from a finite-difference eigenmode solver were used to construct a two-dimensional perturbation coupling matrix that accounts for polarisation dependence and modal asymmetry. Forward and backward optical propagation through a multimode fiber containing six engineered reflection sites was modelled using a continuous spatial transfer-matrix approach, incorporating deformation-induced modal coupling and redistribution of optical power. The resulting spatially resolved back-reflected fields were mapped into the time-of-flight domain, binned at 55 ps intervals, and convolved with the measured SPAD-array instrument response function to reproduce the experimental TCSPC measurement conditions. The simulations reveal progressively stronger spatial and temporal redistribution of the photon-return distributions with increasing fiber deformation (Supplementary Figures S5 \& S6), consistent with the experimental observations. These results provide a physical basis for exploiting deformation-dependent multimode backscattering as an information-bearing mechanism for fiber-shape reconstruction.

\section*{Data availability}
The data that support the findings of this study are openly available in the Heriot-Watt University PURE research data management system.

\providecommand{\noopsort}[1]{}\providecommand{\singleletter}[1]{#1}%

\section*{Acknowledgments}
The authors thank the UK EPSRC for supporting this work via the “U-care” project (EP/T020903/1) and the EPSRC Research and Partnership Hub in Microscale Science and Technology to Accelerate Therapeutic Innovation (MicroTex) (UKRI143).

\section*{Author contributions}

HKC built and characterized the time-of-flight experimental setup and performed the measurements. SNK designed and fabricated the laser-engineered fibers and led the shape reconstruction analysis. SNK led and performed the data processing with inputs from HKC. MAA performed the numerical simulations and was involved in discussions. RRT conceived the project and acquired funding. All authors contributed to interpreting the results and to writing and revising the manuscript. 

\subsection*{Conflict of interest}
HKC, SNK and RRT are named inventors on UK Patent Application No.  2625863.2, relating to some of the technologies and methods described in this paper. The authors declare no other competing interests.

\section*{Additional Information}

See the Supplementary Material for more information.

 \clearpage

\includepdf[
    pages=2,
    pagecommand={
        \begin{center}
            \Large\bfseries Supplementary Information
        \end{center}
    }
]{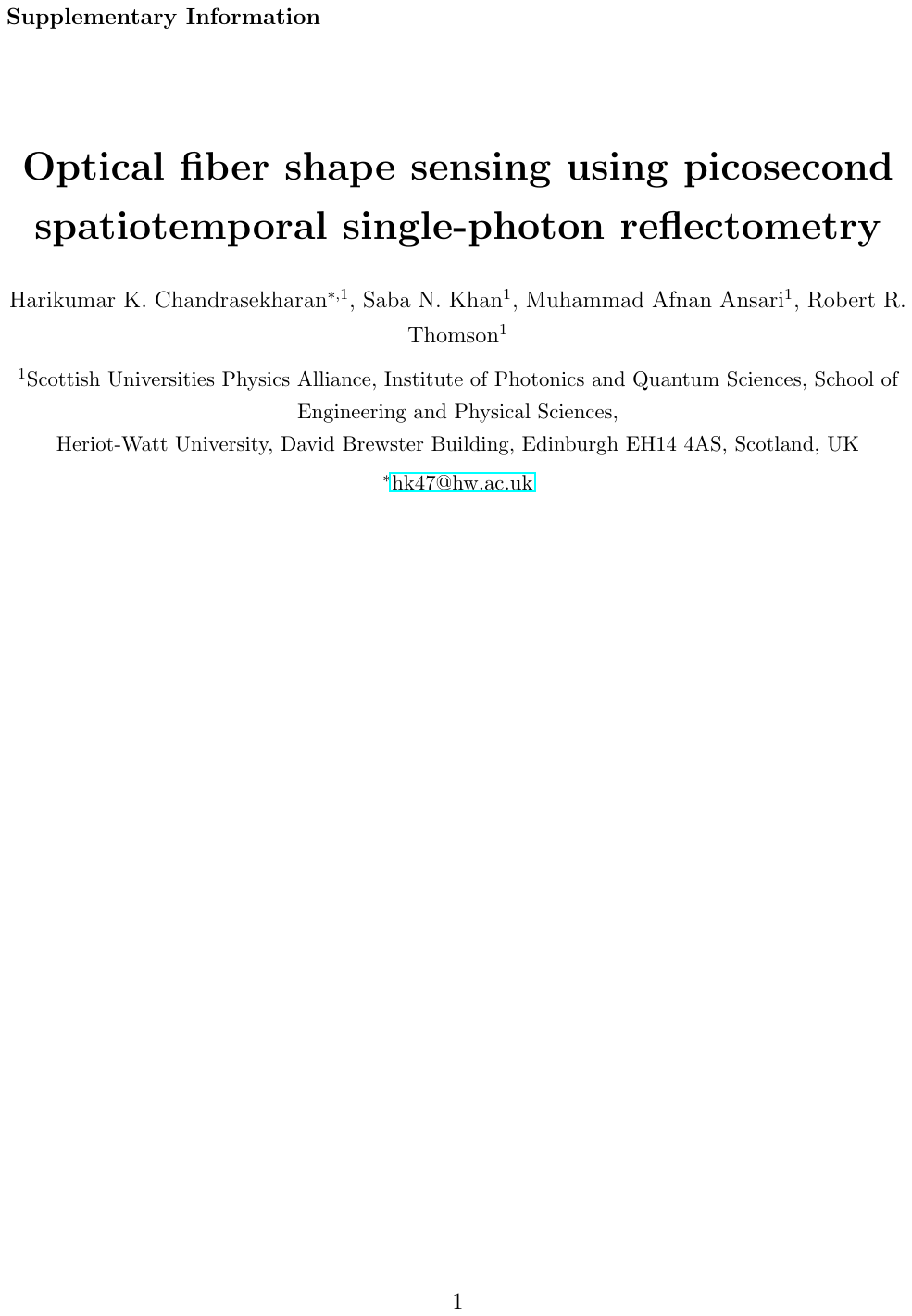}

\includepdf[pages=3-]{AdditionalInfo.pdf}

\end{document}